\documentclass[12pt,twoside]{article}
\usepackage{fleqn,espcrc1}

\title{
\vspace*{-2.8cm}
\hfill{\small { TRI-PP-00-30}}\\
\hfill{\small { June 23, 2000}}\\[1.7cm]
Charged radiative pion capture on the nucleon in heavy baryon 
chiral perturbation theory}

\author{H. W. Fearing\address{TRIUMF, 4004 Wesbrook Mall, Vancouver, 
BC, Canada}, 
Th. R. Hemmert\address{Forschungszentrum J\"ulich, Institut 
f\"ur Kernphysik (Th), J\"ulich, Germany}, 
R. Lewis\address{Department of Physics, University of Regina, 
Regina, SK, Canada}, and
C. Unkmeir\address{Institut f\"ur Kernphysik, 
Johannes Gutenberg-Universit\"at, Mainz, Germany}}

\begin{document}

\maketitle

\begin{abstract}
The differential cross sections and s-wave and p-wave multipoles for $\pi^- p
\rightarrow \gamma n$ and $\pi^+ n \rightarrow \gamma p$ have been calculated
through $O(p^3)$ in heavy baryon chiral perturbation theory (HBChPT).  Fits to
existing data allow several of the low energy constants to be
determined. Generally results of the calculation compare well with dispersion
relation predictions.
\end{abstract}

\section{INTRODUCTION}

Radiative pion capture by the nucleon, via the associated pion photoproduction
process, has been extensively investigated for neutral pions in the context of
heavy baryon chiral perturbation theory (HBChPT). Much less is known however
about the charged pion case, which has contributions from the leading order,
$O(p)$. There the s-wave multipole has been calculated \cite{BKM} and is in
reasonable agreement with experiment. A calculation of the p-wave multipoles
can provide additional interesting information about the convergence of the
HBChPT expansion above absolute threshold and also an opportunity to compare
this approach with others, for example dispersion relation approaches. Such a
calculation is also important as a way to obtain some more of the low energy
constants. The values of such constants are necessary if we are to continue the
program begun for ordinary muon capture in Ref.~\cite{FLMS} and extend it to
radiative muon capture \cite{rmc}.

\section{OUTLINE OF CALCULATION}

We begin with the HBChPT Lagrangian ${\cal L}_{\pi N}={\cal L}_{\pi N}^{(1)}+
{\cal L}_{\pi N}^{(2)}+{\cal L}_{\pi N}^{(3)}$ to $O(p^3)$.  The lowest order
Lagrangian is given by the standard form ${\cal L}_{\pi N}^{(1)} =
\overline{N}_v ( i v\cdot \nabla + g_A S\cdot u) N_v$. The higher order terms
are taken in the specific representation of Ecker and Moj\u{z}i\u{s} \cite{EM}
exactly as used in Ref.~\cite{FLMS}. To $O(p^3)$ it is necessary to calculate
all tree level diagrams together with the one loop diagrams constructed from
the low order parts of the Lagrangian. We performed the calculation in an
arbitrary gauge, so that we could check gauge invariance explicitly, and only
at the end reduced to the specific gauge $v\cdot\epsilon=0$. The calculation
was also performed in a general isospin basis, so that we could verify that the
$\pi^0$ amplitudes agreed with previous work.\cite{pizero}

The reaction of interest here is the radiative pion capture process, whereas
almost all previous work has dealt with the inverse, photoproduction, reaction.
The cross sections are related trivially via the usual detailed balance
equation. However the relations between the amplitudes for the two processes,
and thus between the multipoles, are not so simple and involve complex
conjugation and various sign changes. Thus we have actually evaluated the
amplitude and multipole amplitudes for the photoproduction process, making
connection with the radiative capture data at the cross section level. This
means that we can compare results for amplitudes and multipoles directly with
the conventional ones derived for photoproduction.

With this understanding the amplitude can be written, in terms of the T-matrix,
in the Coulomb gauge with $\epsilon_0=0$ and the transversality condition
$\vec{\epsilon} \cdot \vec{k} = 0$ as \cite{chew,berends}
\begin{eqnarray}\label{gamamp}
{\cal M}^{\gamma N \rightarrow  \pi N} =
\frac{m_N}{4\pi \sqrt{s}} \; T \cdot \epsilon &=&
F_1(E_\pi, x) \; i \chi^\dagger \vec{\sigma} \cdot \vec{\epsilon} \; \chi +
F_2(E_\pi, x) \; \chi^\dagger \vec{\sigma} \cdot \hat{q} \; \vec{\sigma}
\cdot \left( \hat{k} \times \vec{\epsilon} \right) \chi + \nonumber \\
& & F_3(E_\pi, x) \; i \chi^\dagger \vec{\sigma} \cdot \hat{k} \; \vec{
\epsilon} \cdot \hat{q} \; \chi +
F_4(E_\pi, x) \; i \chi^\dagger \vec{\sigma} \cdot \hat{q} \; \vec{\epsilon}
\cdot \hat{q} \; \chi \; , \label{eq:T} \nonumber
\end{eqnarray}
where $\sigma^i$ is a Pauli matrix in spin space between the two-component
spinors of the incoming/outgoing nucleon ($\chi /\chi^\dagger$), $\epsilon$ is
the photon polarization vector and $x=\cos\theta$ corresponds to the cosine of
the angle between the photon and the pion momenta and $E_\pi$ is the pion's
center of mass energy. Furthermore, each structure amplitude $F_i(E_\pi, x)$
($i$=1,2,3,4) can be decomposed into three isospin channels, corresponding,
when the appropriate linear combinations are taken, to the physical
processes. The s-wave electric, p-wave electric and two p-wave magnetic
multipoles can then be projected from these using standard
formulas.\cite{chew,berends}
 
\section{RESULTS}

By squaring the amplitude we obtain expressions for the differential cross
sections which can be fitted to recent data from TRIUMF \cite{Salomon,newdata}
and SAL \cite{Korkmaz} to obtain values for the three unknown low energy
constants, $b_{10}, b_{21}^r, b_{22}^r$, which contribute. The $O(p)$ result
does not fit the data. Adding $O(p^2)$ terms improves the fit, but it is
necessary to include the full $O(p^3)$ result to get a good fit. This is
consistent with applications of HBChPT to other processes which also often find
that the $O(p)$ terms dominate, but $O(p^2)$ and $O(p^3)$ are both important
and comparable corrections. The values $b_{21}^r=-8.2 \pm 0.7$ and $b_{22}^r =
9.2 \pm 0.6$ so obtained are stable and quite reasonable. The constant $b_{10}$
is double valued, based only on the empirical fit to the data. However
comparison to dispersion relation calculations \cite{Tiator}, strongly favors
one of the fits and results in a best value $b_{10} = 13.7 \pm 4.5$.

{}From the amplitudes we can obtain expressions for the s-wave and p-wave
multipoles, in the energy region near threshold and exactly at threshold, in
the two relevant isospin channels, namely, $E_{0+}^{(0,-)}$, $m_{1+}^{(0,-)}$,
$m_{1-}^{(0,-)}$, $e_{1+}^{(0,-)}$. These rather complicated formulas are given
explicitly in Ref.~\cite{HWF}. Then, using the values of the $b_i$ obtained
above and in our earlier calculation of muon capture \cite{FLMS} we can
evaluate these multipoles.  The results for the s-wave multipoles agree well
with previous HBChPT calculations \cite{BKM} and with dispersion analyses
\cite{Tiator}.  The p-wave multipoles are also in reasonable agreement with the
dispersion results for one of the two solutions. The other solution, which has
a very large value of $b_{10}$ also gives multipole amplitudes quite different
than those obtained in the dispersion relation calculation, and it is on that
basis that we rule it out.  In general the convergence is good for the electric
multipoles, whereas for the magnetic multipoles the $O(p^3)$ terms are
comparable to or only a bit smaller than the $O(p^2)$, thus suggesting that it
would be interesting to extend this calculation to $O(p^4)$, which can still be
done within the context of a one loop calculation.

\section{CONCLUSIONS}

To summarize, we have investigated the radiative capture of a charged pion by a
nucleon using heavy baryon chiral perturbation theory and have obtained
explicit expressions for the amplitude and for the s- and p-wave
multipoles. Fits to the available cross section data allowed us to obtain the
three necessary low energy constants. Using the values so obtained, the eight
s- and p-wave multipoles (four for the $\pi^+$ case and four for the $\pi^-$
case) were calculated and compared with results previously obtained from
dispersion theory \cite{Tiator}. In general the agreement was good, though the
convergence of the results for the magnetic multipoles was not as fast as for
the electric ones, thus suggesting that it might be valuable to consider
extending the present work to $O(p^4)$ or to include explicit $\Delta(1232)$
fields in the chiral Lagrangian.

This work was supported in part by the Natural Sciences and Engineering
Research Council of Canada.

\end{document}